\documentclass[12pt]{article}
\usepackage{graphicx, epsfig}
\def\be{\begin{equation}}
\def\ee{\end{equation}}
\def\ba{\begin{eqnarray}}
\def\ea{\end{eqnarray}}
\def\la{~\mbox{\raisebox{-.6ex}{$\stackrel{<}{\sim}$}}~}

\def\bq{\begin{quote}}
\def\eq{\end{quote}}

 at 10truept

\newcommand{\beq}{\begin{equation}}
\newcommand{\eeq}{\end{equation}}
\newcommand{\beqa}{\begin{eqnarray}}
\newcommand{\eeqa}{\end{eqnarray}}

\def\la{~\mbox{\raisebox{-.6ex}{$\stackrel{<}{\sim}$}}~}

\def\la{~\mbox{\raisebox{-.6ex}{$\stackrel{<}{\sim}$}}~}

\def\B{{\bf B}}

\def\X{{\Xi}}
\def\h{s_*}
\def\B{{\cal B}}

\def\bk{{\bf k}}

\def\beq{\begin{equation}}
\def\eeq{\end{equation}}
\def\P{{\cal P}}
\def\Pinf{\P_{\rm inf}}
\def\Ptrap{\P_{\rm trap}}
\def\N{{\cal N}}

\def\m{\sigma} 
\def\mod{\sigma}

\def\ltap{\ \raise.3ex\hbox{$<$\kern-.75em\lower1ex\hbox{$\sim$}}\ }
\def\gtap{\ \raise.3ex\hbox{$>$\kern-.75em\lower1ex\hbox{$\sim$}}\ }
\def\gl{\ \raise.5ex\hbox{$>$}\kern-.8em\lower.5ex\hbox{$<$}\ }
\def\roughly#1{\raise.3ex\hbox{$#1$\kern-.75em\lower1ex\hbox{$\sim$}}}

\begin{document}

\thispagestyle{empty}
\begin{flushright}
arXiv:\ 0906.1813\\ June 2009
\end{flushright}
\vspace*{1cm}
\begin{center}
{\Large \bf Primordial perturbations and non-Gaussianities\\
\vspace{0.3cm}
from  modulated trapping}\\
\vspace*{1.5cm} {\large David Langlois$^{\dagger,\ddagger}$\footnote{\tt
langlois@apc.univ-paris7.fr} and
Lorenzo Sorbo$^{*,}$\footnote{\tt sorbo@physics.umass.edu}}\\
\vspace{.5cm} {\em $^{\dagger}$ APC (Astroparticules et Cosmologie), UMR 7164 (CNRS, Universit\'e Paris 7), \\
10, rue Alice Domon et L\'eonie Duquet,
 75205 Paris Cedex 13, France}\\
\vspace{.15cm} {\em $^{\ddagger}$ Institut d'Astrophysique de Paris (IAP), 98bis Boulevard Arago, 75014 Paris, France}\\
\vspace{.15cm} {\em $^{*}$ Department of Physics,
University of Massachusetts, Amherst, MA 01003, USA}\\
\vspace{.15cm} \vspace{1.5cm} ABSTRACT
\end{center}

We propose a new mechanism to generate primordial curvature  perturbations, based on the resonant production of particles during inflation. It is known that this phenomenon  can trap the inflaton for a fraction of e-fold.
This effect  is governed by the mass of the produced particles and by their coupling to the inflaton, parameters which can depend on the expectation value of other fields. If   one of such additional fields -- a {\em modulaton} -- is light, then its fluctuations, acquired during the earlier stages of inflation, will induce a spatial modulation of the trapping, and thus of the end of inflation, corresponding  to a curvature perturbation.
We calculate the power spectrum, bispectrum and  trispectrum of the curvature perturbations generated by this mechanism, taking into account the perturbations due to the inflaton fluctuations as well. We find that modulated trapping could provide the main contribution to the observed power spectrum and lead to detectable primordial non-gaussianities.

\vskip2.5cm

\begin{flushleft}
\end{flushleft}


\setcounter{footnote}{0}

\section{Introduction}

Soon after being proposed as a mechanism to overcome the shortcomings of the standard hot Big Bang scenario, inflation turned out to come with a very rich bonus: it provides the seed inhomogeneities that eventually will evolve to form the structures we inhabit (see e.g.~\cite{Mukhanov:2005sc}). In this standard picture, there is one effective degree of freedom, the inflaton, that fulfills the double task of driving inflation (via its zero mode) and of generating the primordial spectrum of perturbations (via its quantum fluctuations). 

More recently, it has been realized that these two purposes can be fulfilled by different fields and  one can envisage scenarios where primordial perturbations are initially stored in the fluctuations of a second, initially subdominant, scalar field and subsequently transferred to the dominant matter component of the Universe. In more technical terms, this corresponds to a transfer of an initial isocurvature, or entropy, mode into a final curvature, or adiabatic mode. 
 This transfer can occur {\em (i)} during inflation in multi-inflaton scenarios; {\em (ii)} just at the end of inflation, like in the modulated reheating scenario \cite{modulated}, where the subdominant scalar field controls the decay of the inflaton into ordinary matter;  {\em (iii)} long after the end of inflation, like in the curvaton scenario \cite{curvaton}, where the initially subdominant curvaton field comes to dominate after reheating as it redshifts more slowly than radiation. 

Another aspect of inflation which has been actively investigated in the last few years is the resonant production  of particles, which can arise  due to the coupling of the inflaton to other fields.  Indeed, if the inflaton is coupled to a field $\chi$, either bosonic or fermionic, then $\chi$ particles can be produced by resonant effects whenever their effective mass crosses zero as the inflaton evolves.
The most studied  example is preheating~\cite{Kofman:1997yn} (see also~\cite{Dolgov:1989us}), where particle production occurs while the inflaton is oscillating at the bottom of its potential. However, this effect can also take place {\it during inflation} as pointed out in \cite{ckrt}. In the latter case, the backreaction of the produced particles
induces a slow down  of the inflaton $\phi$, effectively trapping it for a fraction of e-fold before the quanta of $\chi$ are diluted away by the expansion of the Universe. 
One consequence of this slow-down of the inflaton is the generation of features in the inflationary spectrum~\cite{ckrt,Elgaroy:2003hp,Romano:2008rr}. The same mechanism can also help halt moduli at points with enhanced symmetry~\cite{Kofman:2004yc}. More recently, it has been shown in~\cite{Green:2009ds} that repeated trapping events can slow down the inflaton enough to lead to slow-roll inflation even in the presence of a steep inflationary potential.

In the present paper, we consider a
scenario where the intensity of the trapping  depends on an additional light scalar field, $\m$, which we will  call a `modulaton' (this field does not contribute significantly to the energy density, neither during inflation, like in multi-field inflation,  or after inflation, like in the curvaton case). More precisely, we will assume that the mass of $\chi$, or its coupling to the inflaton, depends on the modulaton $\m$, which is light during inflation and thus acquires a quasi scale-invariant spectrum of super-Hubble fluctuations. 
Consequently,  in super-Hubble regions of the Universe with different values of $\m$, inflaton trapping  will occur at different times and/or  will be more or less strong. 
 The duration of inflation will thus vary from one region to another.  
This way the perturbation in the $\chi$ modes will be transferred into a curvature perturbation. 

This idea of using a modulaton is not new. This is indeed the key ingredient to the inhomogeneous or modulated reheating scenario \cite{modulated}, where the decay rate of the inflaton depends on the modulaton $\m$. It was extended recently \cite{Kohri:2009ac} to modulated preheating, where a phase of preheating
takes place instead of 
perturbative reheating.
In this
case, the modulaton enters into the coupling $g$ between the inflaton and the produced particles and the fluctuations of the modulaton are transferred into curvature perturbations, because both the duration of the preheating phase and the density energy of the scalar field at the end of preheating depend on $g$ (similar ideas are discussed in \cite{Matsuda:2007tr}). 

While using the ideas of modulaton and particle production, our scenario is very different from the previous ones because the crucial effect here is  the {\it backreaction} of the particle production on the  motion  of the inflaton or, more generally, of any scalar field dominating the energy density in the Universe. After they have temporarily  trapped the inflaton, the produced particles become cosmologically irrelevant as they are rapidly diluted by expansion.

Using the $\delta N$ formalism, we compute the amplitude of the curvature perturbations generated by modulated trapping and we compare their contribution in the power spectrum with the contribution due to the usual inflaton perturbations. We distinguish the case where the coupling 
depends directly on the modulaton from the case where the modulaton affects only the critical value of the inflaton at which particle production takes place. In the latter case, the contribution from modulated trapping is usually small in the simplest models of inflation but it can become dominant if the 
Hubble parameter at particle production is much smaller than when the modulaton fluctuations were generated. 
In the former case, modulated trapping can easily dominate.
We also investigate the primordial non-Gaussianities generated by this mechanism. More precisely, we compute the bispectrum and the trispectrum and we find that modulated trapping can lead to a level of nongaussianities that would be detectable with future experiments. 

The plan of our paper is as follows. In section 2 we present the model and review the main results concerning the impact of 
resonant particle production on the evolution of the slow-rolling inflaton. 
In section 3 we derive the linear perturbations generated by modulated trapping and compare their contribution with that due to the inflaton fluctuations, for different cases. In section 4 we investigate  non-Gaussianities by computing the bispectrum and trispectrum. 
Finally, we conclude in section 5.

\section{Model and homogeneous solution}

We consider a model where the inflaton $\phi$  is coupled to other fields $\chi$, which can be either bosonic or fermionic. In addition to the usual kinetic terms and the self-interaction  potential $V\left(\phi\right)$ of the inflaton, we assume that the Lagrangian contains, in the bosonic case, a term of the form
\beq
{\cal L}_{\rm int}=-\frac{1}{2} {\cal N}\left(m-\lambda\phi\right)^2\chi^2,
\eeq
and, in the fermionic case,
\beq
{\cal L}_{\rm int}=-{\N}\left(m-\lambda\phi\right)\,\bar{\chi}\chi,
\eeq
In both cases,  $\cal{N}$ denotes the number of species of particles with the same mass $m$ and the same coupling $\lambda$.
Taking into account the coupling to the inflaton, these particles have an effective mass 
\beq
M\left(\phi\right)=m-\lambda\phi,
\eeq
 so that there is a critical value of $\phi$, denoted by 
 \beq
 \phi_*\equiv m/\lambda
 \eeq
  where $M\left(\phi\right)$ vanishes. As discussed in~\cite{Kofman:1997yn}, resonant particle production will occur when $\phi$ crosses $\phi_*$. In this section we will review the main results of \cite{ckrt}, where resonant production (of fermions) during inflation was first studied. As in~\cite{ckrt}, we will focus on fermion production, keeping in mind that the scenario where scalars are resonantly produced gives the same results.
  
When $M\left(\phi\right)$ crosses zero, the particle occupation number suddenly increases from zero to the value
\beq
n_*=\frac{\lambda^{3/2}}{2\pi^3}v_*^{3/2}, \qquad v_*\equiv |\dot\phi_*|\, ,
\eeq
where we use the subscript $*$ to denote the time of particle production. 
The number of particles is then diluted by the expansion so that 
\beq
n(t)=n_*\left(\frac{a}{a_*}\right)^{-3}\Theta(t-t_*)\, ,
\eeq
where $\Theta$ is the Heaviside distribution.
The backreaction of the particle production can be estimated by using the Hartree approximation in the equation of motion of the inflaton, which after substituting the particle number obtained above, becomes 
\begin{equation}\label{kg}
\ddot{\phi}+3\,H\,\dot\phi+V'\left(\phi\right)=\N\,\lambda\langle \bar\chi\chi\rangle =\lambda \N n_*\left(\frac{a}{a_*}\right)^{-3}\Theta(t-t_*)
\end{equation}

Since the production and subsequent dilution of the particles occurs during a fraction of e-fold, we will  assume $H=H_*$, $V'\left(\phi\right)=V'\left(\phi_*\right)$ to be constant during the entire process. One can then easily integrate the equation of motion for $\phi$. Denoting 
\beq
\Delta\phi\left(t\right)\equiv\phi(t,\,\lambda\neq 0)-\phi(t,\,\lambda=0)
\eeq
the difference between the solutions with and without particle production, one finds
\begin{equation}\label{deltaphidot}
\Delta\dot\phi\left(t>t_*\right)=\N\,\lambda \,n_*\, e^{-3H_*\left(t-t_*\right)}\,\left(t-t_*\right)\,\,.
\end{equation}
We have assumed here implicitly that $\dot\phi<0$ and the positive $\Delta\dot\phi$ thus corresponds to a decrease of the absolute value of the inflaton velocity. Since particle production occurs at the expense of the kinetic energy of the inflaton, the sign of $\Delta\dot\phi$ will always be opposite to that of $\dot\phi$, so that $|\dot\phi+\Delta\dot\phi|<|\dot\phi|$.

This slow-down of the inflaton, 
illustrated in Fig.~1, leads to the generation of features in the inflationary spectrum since its amplitude  depends on $\dot\phi$. This has first been 
pointed out in \cite{ckrt} and subsequently studied in \cite{Elgaroy:2003hp} and \cite{Romano:2008rr}. A scenario where the resonant production of particles during inflation leads to features in the observable curvature spectrum has also been discussed in \cite{Langlois:2004px}. In the present work, however, we will assume that particle production occurs much later than when the scales of cosmological interest crossed out the Hubble radius. Consequently, the features in the spectrum will affect scales which are much smaller than those
corresponding to the present cosmological window. This also applies to the fluctuations generated by the re-scattering of the produced particles with the homogeneous scalar field, which have been recently studied  in~\cite{Barnaby:2009mc}.

\begin{figure}[h]
\centering
\includegraphics[width=0.5\textwidth, clip=true]{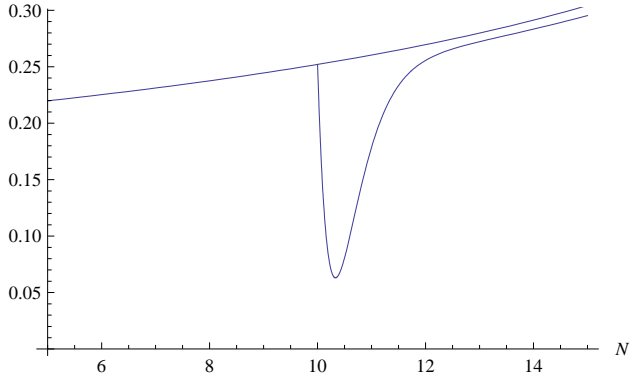}
\caption{Evolution of $|\dot\phi|$ as a function of the number of e-folds $N$ for a model of chaotic inflation with trapping (lower curve) and without trapping (upper curve).}
\end{figure}

Of course, the energy injected into the produced particles cannot exceed the kinetic energy of the inflaton before production, and the value of $\Delta\dot\phi$ is limited by the fact that the rolling of $\phi$ will at most be halted by trapping. In other words Max$\vert\Delta\dot\phi\vert$ cannot exceed 
$v_*\equiv \vert\dot\phi_*\vert$. The behavior of the system when backreaction effects are important has been studied numerically in~\cite{ckrt}, where it was found that the evolution of $\Delta\dot\phi$ has still the functional form of~(\ref{deltaphidot}), but with a different overall normalization that ensures energy conservation. We can account for a such a behavior by defining the parameter 
\beq
\label{beta}
\beta\equiv {\rm Max}(\Delta\dot\phi)/v_*=\frac{\N\,\lambda \,n_*}{3H_* e \, v_*}  =\frac{\N \, \lambda^{5/2}\,  v_*^{1/2}}{6\pi^3\, e\, H_*} \, ,
\eeq
which must satisfy $\beta\la 1$. The above expression can be used to trade $\lambda$ for $\beta$. It is also useful to note that, if the inflaton is in slow-roll, its velocity $v_*$ just before particle production is given by
\beq
\label{v}
v_*=\sqrt{2\epsilon_*}\,H_*M_P\, ,
\eeq
where $\epsilon$ is the first  of the usual slow roll parameters 
\beq
\label{SR_parameters}
\epsilon\equiv\frac12\left(\frac{M_PV'}{V}\right)^2\, , \qquad \eta\equiv \frac{M_P^2V''}{V}\, .
\eeq 
and $M_P\equiv (8\pi G)^{-1/2}$ is the reduced Planck mass.

\begin{figure}[h]
\centering
\includegraphics[width=0.5\textwidth, clip=true]{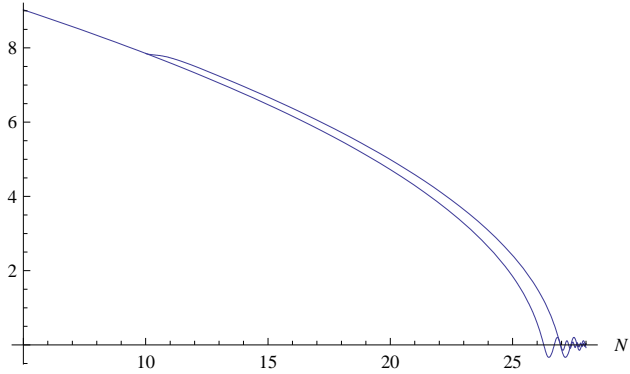}
\caption{Evolution of $\phi$ as a function of the number of e-folds $N$ with trapping (upper curve) and without trapping (lower curve), using the same model and initial conditions as in Fig.~1.}
\label{phi}
\end{figure}

The inflaton is only momentarily slowed down, as the term on the right hand side of (\ref{deltaphidot}) quickly goes to zero
after the production has occurred ($t-t_*\gg 1/H_*$). However, the inflaton will have  accumulated a delay with respect to its evolution without particle production:
\begin{equation}\label{delphi}
\Delta\phi=\int_{t_*}^\infty \Delta\dot\phi\,dt
=\frac{\N\lambda n_*}{9H_*^2}
=\frac{\lambda^{5/2}\N v_*^{3/2}}{18\pi^3 H_*^2}
\end{equation}
Consequently, inflation with particle production will end later than inflation without particle production, as one can see clearly on Figure~\ref{phi}.

\section{Fluctuations}

We now assume that the fermions (or bosons) are also coupled to another light scalar field $\mod$, which we will name the {\em modulaton} field to distinguish it from the inflaton, so that the effective mass  
\beq
m_{\rm eff}=m(\mod)-\lambda(\mod)\,\phi\, 
\eeq
depends on $\mod$. One possibility, which we will consider later as a specific example, is that the  coupling to $\mod$ arises from a Yukawa coupling $-g\,\mod\,\bar\chi\chi$, which implies $m(\mod)=g\mod$. 

The critical value for $\phi$ where particle production occurs now depends  on 
 $\mod$ and is given by  
\beq
\phi_*(\mod)=\frac{m(\mod)}{\lambda(\mod)}.
\eeq
Since the modulaton $\mod$ is assumed to be light, i.e. its mass is much smaller than the Hubble parameter during inflation, it acquires some fluctuations on super-Hubble scales, via amplification of its quantum fluctuations. 
As a consequence, the value of the scalar field $\mod$ fluctuates from one patch of the Universe to another, and therefore particle production is slightly different in each of these patches. The 
time delay in the evolution of the inflaton, calculated in the previous section, will thus fluctuate from one patch to the other,  which can be reinterpreted as fluctuations in the final curvature perturbation at the end of inflation. 

In order to quantify the curvature perturbations generated by the  fluctuations of $\mod$, it is convenient to use the so-called $\delta N$ formalism \cite{deltaN}, based on the local number of e-folds, or integrated expansion, between some initial and final hypersurfaces
\beq
N(x)=\int_i^f H(t,x)\,dt \, .
\eeq
In particular, the curvature perturbation on a uniform energy density final hypersurface, which we will denote $\zeta$, is directly related to the perturbation of the number of e-folds defined with respect to some initial flat hypersurface, 
\beq
\zeta=\delta N\equiv N(x)-\bar{N},
\eeq
where $\bar{N}$ is the number of e-folds in the homogeneous background spacetime. 
This is true not only at linear order but also at non-linear order \cite{Lyth:2004gb,Langlois:2005ii,Langlois:2005qp}, which will enable us to compute the non-Gaussianities in the next section.

During slow-roll inflation, the number of e-folds is 
\beq
N_{\rm slow-roll}= \int_{\phi_i}^{\phi_f}\frac{H}{\dot\phi}\,d\phi=-\frac{1}{M_P^2}\int_{\phi_i}^{\phi_f}\frac{V}{V'}\,d\phi\, ,
\eeq
where we have used the slow-roll equations $\dot\phi=-V'/(3H)$ and $3H^2=V/M_P^2$ to obtain the second equality.
In our case, the above expression is not valid during the bursts of fermion production, where the scalar field deviates from the slow-roll solution. This induces a net change in the number of e-folds,
\beq
\label{DeltaN}
\Delta N=-H_*\,\frac{\Delta\phi}{\dot\phi_*}=\frac{\N\lambda n_*}{9H_* v_*}
=\frac{\lambda^{5/2}\N v_*^{1/2}}{18\pi^3 H_*}
\eeq
where we have used (\ref{delphi}). 
This shift in the number of e-folds depends on the scalar field $\mod$: either directly in $\lambda$, or indirectly via the time $t_*$ of particle production, which depends on the critical value $\phi_*(\mod)$.

\subsection{The power spectrum}

Since the masses of both the inflaton $\phi$ and (by assumption) the modulaton $\mod$  are smaller than the Hubble parameter during inflation, both $\phi$ and $\mod$ will carry perturbations on super-Hubble  scales, characterized by the power spectra 
\beq
\label{fluctuations}
\P_{\delta\phi}=\P_{\delta\mod}=\left(\frac {H_k}{2\pi}\right)^2\, ,
\eeq
where the index $k$ denotes the Hubble crossing ($k=aH_k$), for the cosmological scales of interest. 

By expanding the number of e-folds at first order as a function of the perturbations of the inflaton, $\delta \phi$, and of the 
modulaton, $\delta\mod$, one can write the curvature perturbation 
 $\zeta$ as
\beq
{\cal \zeta}=\delta N=\delta N_{\rm slow-roll}+\delta \Delta N=-\frac{1}{M_P^2}\frac{V_k}{V_k'}\,\delta\phi
+\Delta N_{,\mod}\,\delta\mod\, .
\eeq
This implies, since the perturbations in $\phi$ and in $\mod$ are uncorrelated, that the  curvature power spectrum is given by the sum of two contributions,
\beq\label{power}
{\cal P}_\zeta\equiv\Pinf+\Ptrap=
\frac{V_k^2}{M_P^4 {V_k'}^2}{\cal P}_{\delta\phi}+(\Delta N_{,\mod})^2\P_{\delta\mod}=  \left[\frac{1}{2\epsilon_k}+(\Delta N_{,\mod})^2 M_P^2\right]\,\left(\frac{H_k}{2\pi\,M_P}\right)^2\, ,
\eeq
where we have used (\ref{SR_parameters}) and (\ref{fluctuations})  in the last equality.

The importance of the trapping effect with respect to the usual inflaton fluctuation term can be quantified by 
 its relative contribution in the total power spectrum, defined by 
\beq
\X\equiv \frac{\Ptrap}{\P_\zeta},
\eeq
so that 
\beq
\label{X}
\frac{\X}{1-\X}=\frac{\Ptrap}{\Pinf}=2\epsilon_k\,(\Delta N_{,\mod})^2 M_P^2\, .
\eeq
One recovers the standard inflation result when $\X\ll 1$. The opposite limit is $\X= 1$ where the trapping dominates. The intermediate values of $\X$ correspond to situations where both terms can  significantly contribute. This is quite  similar to mixed inflaton and curvaton models, investigated 
in~\cite{Langlois:2004nn},
where both the inflaton and the curvaton fluctuations contribute to the observed spectrum. 

The spectral index is also affected by the fact that the two contributions have different scale dependence, since the inflaton contribution is proportional to $H_k^2/\epsilon_k$ while the trapping contribution is simply proportional to $H_k^2$. One finds that the spectral index is given, in terms of the slow-roll parameters and of the trapping fraction $\X$, by the expression
\beq
n_s-1\equiv \frac{d \ln \P_\zeta}{d\ln k}=2(1-\X)\, \eta_k+(4\X-6)\, \epsilon_k,
\eeq
at leading order in the slow-roll parameters. It can be checked that this result is exactly the same as in the mixed inflaton-curvation scenario~\cite{Langlois:2004nn}. If the inflaton contribution dominates, one recovers the standard result $n_s-1=2\eta_k-6\epsilon_k$ while one finds 
$n_s-1=-2\epsilon_k$ if modulated trapping dominates, like in  the pure curvaton scenario. 

Let us now compute $(\Delta N_{,\mod})^2$
explicitly. Substituting the expression of $v_*$ in terms of $\epsilon_*$ and $H_*$ into 
(\ref{DeltaN}) yields 
\beq
\label{DeltaN2}
\Delta N=\frac{\N\,\lambda^{5/2}}{18\,\pi^3}\,\frac{M_P^{1/2}}{H_*^{1/2}}\,\left(2\epsilon_*\right)^{1/4}\,,
\eeq
where $\epsilon_*=\epsilon(\phi_*)$ and $H_*=H(\phi_*)$ depend on $\mod$ via $\phi_*(\sigma)=m/\lambda$.
Using
\beq
\frac{\epsilon_{,\phi}}{\epsilon}= 2 \left(\frac{V''}{V'}-\frac{V'}{V}\right)=\frac{\sqrt{2}}{M_P\sqrt{\epsilon}}\left(\eta-2\epsilon\right), \qquad
 \frac{H_{,\phi}}{H}=\frac{V'}{2V}=\frac{\sqrt{2\epsilon}}{2 M_P}
\eeq
and 
\beq
\frac{d\phi_*}{d\mod}=\frac{d}{d\sigma}\left(\frac{m}{\lambda}\right)\, ,
\eeq
one finds
\beq
\label{DeltaNprime}
\frac{\Delta N_{,\mod}}{\Delta N}=\frac52\frac{\lambda'}{\lambda}+\frac{1}{4M_P}\sqrt{\frac{2}{\epsilon_*}}\left(\eta_*- 3 \epsilon_*\right)
\left(\frac{m}{\lambda}\right)'\,,
\eeq
where the primes in the above formula denote a derivative with respect to $\mod$.
We thus finally obtain for the power spectrum due to modulated trapping
\beq
\label{P_trap}
\Ptrap=
(\Delta N_{,\mod})^2\,\left(\frac{H_k}{2\pi}\right)^2
=
\frac{\N^2\,\lambda  \left[-5\,M_P \sqrt{2\,\epsilon_*}\,\lambda\,\lambda '+ \left(\eta_* -3\epsilon_*\right)\,\left(m\,\lambda '-\lambda \,m'\right)\right]^2}{1296\,\pi^6\, M_P H_* \sqrt{2\,\epsilon_*}}\,\left(\frac{H_k}{2\pi}\right)^2\,.
\eeq

In the next two subsections, we will consider two particular cases of the above general formula. In both cases, we will assume that the mass of the fermions depends linearly on $\mod$, i.e. $m=g\,\mod$ where $g$ is constant, as would result from a standard Yukawa coupling between $\mod$ and the fermions. 
We will moreover assume that   the coupling $\lambda$ is independent of $\mod$ in the first case,  whereas it is of the form 
$\lambda=\sigma/M$, such as would arise from a non-renormalizable term, in the second case. In the latter case, $\lambda$ and $m$ have the same linear dependence on $\mod$ and therefore $\phi_*$ becomes independent of $\mod$. 
Of course, more general cases, combining both direct and indirect dependence on $\mod$ can be envisaged, and the purpose of our two examples is simply to analyze separately the two types of dependence on $\mod$.

\subsection{Modulaton-independent coupling}

We now assume 
\beq
\lambda=\lambda_0, \qquad m=g\, \mod,
\eeq
where $\lambda_0$ and $g$ are constants. The trapping effect then depends indirectly on $\mod$ via the critical value $\phi_*(\mod)$.

Then, the expression (\ref{P_trap}) reduces  to
\beq
\label{P_trap1}
\Ptrap=
\left(\frac{1}{72\,\pi^4}\right)^2g^2\,\N^2\, \lambda_0^3\,\frac{(\eta_* -3\,\epsilon_*)^2}{\sqrt{2\,\epsilon_*}}\frac{H_k^2}{H_*\, M_P}.
\eeq
Reexpressing $\lambda_0$ in terms of the braking parameter $\beta$, via (\ref{beta}), the corresponding amplitude can be written in the form
\beq
\Ptrap^{1/2}=\Lambda_*f_{*k}^{-1/5} \left(\frac{H_k}{M_P}\right)^{4/5},
\eeq
where we have introduced the fraction $f_{*k}= H_*/H_k$ and
\beq
\label{Lambda1}
\Lambda_*
= \left(\frac{e^3}{12^7\
\pi^{11}}\right)^{1/5} g \, {\cal N}^{2/5}\, \beta^{3/5}\,  \h 
\simeq 4.5\times 10^{-3} g \, {\cal N}^{2/5}\, \beta^{3/5}\,  \h \, ,\qquad 
\h(\epsilon_*, \eta_*)\equiv \epsilon_*^{-2/5}\left|\eta_*-3\,\epsilon_*\right|\, .
\eeq

Let us now see whether the trapping contribution can dominate the power spectrum, i.e. $\X\simeq 1$. If this is the case, then the  trapping 
amplitude (\ref{P_trap1}) must agree 
 with the observed amplitude of the primordial spectrum of perturbations
\begin{equation}
\P_{\zeta, \rm obs}^{1/2}
 \simeq 4.9\times 10^{-5}\ .
 \label{cobe}
\end{equation}
This implies the condition 
\beq
\label{Hk}
\frac{H_k}{M_P}\simeq 4.1\times 10^{-6}\ \Lambda_*^{-5/4} f_{*k}^{1/4}\, .
\eeq
Moreover, we must check that the inflaton contribution is much smaller than the trapping contribution, which implies the condition
\beq
\label{domination}
\frac{\X}{1-\X}=8\pi^2\epsilon_k\Lambda_*^2\left(\frac{H_*}{M_P}\right)^{-2/5} \gg 1\, .
\eeq
Replacing $\Lambda_*$ by  (\ref{Lambda1}) and $H_k$ (\ref{Hk}) thus yields the condition 
\beq
\label{domination1}
g^{5/2}\,  {\N}\,  \beta^{3/2}\,f_{*k}^{-1/2}\,  \epsilon_k\, \h^{5/2}\gg 64
\eeq
on the various parameters of the model, for a domination of the trapping effect. 

If one takes, for instance, $g=1$, $\N=100$ (the same choice of parameters was considered in~\cite{ckrt}) with $\lambda_0$ large enough to make backreaction important ($\beta\simeq 0.8$), one finds that $g^{5/2}\,  {\N}\,  \beta^{3/2}\simeq 71$. Since the slow-roll parameter $\epsilon_k$ must be much smaller than one, this means that the condition (\ref{domination1}) cannot be satisfied unless 
$\epsilon_*\ll \eta_*$ or if the Hubble parameter varies very much between the time when  the perturbations are generated and the time when trapping occurs, so that  $f_{*k}\ll 1$. The former case, i.e.   $\epsilon_*\ll \eta_*$, can arise if inflation at the time of particle production behaves like in hybrid inflation. The latter case can occur if inflation proceeds through different phases with a hierarchy between the various energy scales (like in double inflation), or even if the Universe underwent several distinct phases of inflation, as  discussed for instance in \cite{Adams:1997de} or \cite{Burgess:2005sb}.

Even if the trapping is not the main contribution, it can nevertheless represent a non-negligible fraction of the overall fluctuation spectrum, when
 the parameters $\epsilon_k$ and $\h$ are not too small. For instance, in the case where $\epsilon_*\simeq \eta_*\simeq 0.5$,  $\epsilon_k\simeq 0.01$ and $f_{*k}\simeq 0.1$, one obtains $\Xi\simeq  0.1$.

\subsection{Modulaton-dependent coupling}

We now assume
\beq
\label{lambda_m_2}
\lambda=\frac{\mod}{M}, \qquad m=g\, \mod,
\eeq
where  $g$ is constant and where $M$ is some mass scale. In this case, $\Delta N$ depends on $\mod$ directly via the coupling $\lambda$, while the critical value $\phi_*$ of the inflaton at which trapping occurs is independent of $\mod$. Substituting (\ref{lambda_m_2}) in (\ref{P_trap}), we obtain
\begin{equation}
\label{P_trap2}
\Ptrap=\left(\frac{5}{72\,\pi^4}\right)^2\,{\cal {N}}^2\,\lambda^3\,\sqrt{2\,\epsilon_*}\ \frac{H_k^2\, M_P}{H_*\, M^2}\,\,.
\end{equation}
By comparing this expression with the analogous one of the previous subsection, i.e. (\ref{P_trap1}), we see that the effect of modulated trapping can be much stronger in the present case. The numerical factor $25$, the different dependence on the slow roll parameters, and especially the fact that $M_P/M$ can be much larger than one, 
all concur to increase the contribution of the trapping and therefore push $\Xi$ towards unity.

As in the previous subsection, we can  write the amplitude of perturbations due to trapping in the form 
\beq
\Ptrap^{1/2}=\Lambda_*\, f_{*k}^{-1/5}\,  \left(\frac{H_k}{M_P}\right)^{4/5},
\eeq
with
now
\begin{eqnarray}
\Lambda_*=5\sqrt{2}\left(\frac{e^3}{12^7\,\pi^{11}}\right)^{1/5}
\,\beta^{3/5}\, {\cal{N}}^{2/5}\, \epsilon_*^{1/10}\,  
\frac{M_P}{M}
\ \simeq \ 
3\times 10^{-2}\,\beta^{3/5}\, {\cal{N}}^{2/5}\, \epsilon_*^{1/10}
\,  \frac{M_P}{M}\,\, .
\end{eqnarray}
This gives for example $\Ptrap^{1/2}\simeq 0.2\,  f_{*k}^{-1/5} \,M_P^{1/5}\,H_k^{4/5}/{M}$ when we take  $\epsilon_*=0.5$, ${\cal {N}}=100$ and $\beta=0.8$, as in the previous subsection.

The contribution due to the trapping dominates the observed spectrum if the condition (\ref{domination}) is satisfied.
 Using the present expression for $\Lambda_*$, this condition can be rewritten as
\beq
\label{domination2}
  {\N}\,  \beta^{3/2}\,f_{*k}^{-1/2}\,  \epsilon_k\, \epsilon_*^{1/4} \left(\frac{M}{M_P}\right)^{-5/2}\gg 0.5\, ,
\eeq
 which is rather easily realized, either with a large number of species or a small ratio $M/M_P$. 

\section{Non-Gaussianities}

Let us now study the non-Gaussianities in the modulated trapping scenario. In standard single field inflation, primordial non-Gaussianities are negligible but in many other scenarios, they could be significant. 
 An easy way to compute the non-Gaussianities of the curvature perturbation is to use the Taylor  expansion of the local number of e-folds
  in terms of the scalar field fluctuations \cite{Lyth:2005fi}. 
For a multi-field system, this gives 
\beq
\zeta=\delta N= N_{I} \, \delta \varphi^I + \frac{1}{2}
 N_{IJ} \, \delta \varphi^I \, \delta \varphi^J+\frac16 N_{IJK} \, \delta \varphi^I \, \delta \varphi^J \,\delta\varphi^K+\dots,
\eeq
where we use the implicit summation convention for the field indices $I, J, K, \dots$ and the notation
 $N_{I}\equiv \partial N/\partial \varphi^I$, $N_{IJ}\equiv \partial^2 N/\partial\varphi^I\partial\varphi^J$, etc. 
 In Fourier space, 
this implies that the three-point function is given by
\begin{eqnarray}
\label{bispectrum}
\langle \zeta_{\bk_1} \zeta_{\bk_2} \zeta_{\bk_3} \rangle &\equiv& (2 \pi)^3
\delta^{(3)}(\sum_i \bk_i) 
B_\zeta (\bk_1,\bk_2,\bk_3)
=
 N_{I} N_{J} N_{K} \langle \delta \varphi^I_{\bk_1}
\delta \varphi^J_{\bk_2} \delta \varphi^K_{\bk_3}\rangle + \nonumber
\\ && + \frac{1}{2}  N_{I} N_{J} N_{KL} \langle \delta
\varphi^I_{\bk_1} \delta \varphi^J_{\bk_2} (\delta \varphi^K \star
\delta \varphi^L)_{\bk_3}\rangle
+{\rm perms}, 
\end{eqnarray}
where the symbol $\star$ denotes a convolution product. 

If the scalar field fluctuations are independent and Gaussian, which is expected to be a good approximation for  inflation with standard kinetic terms and in the slow-roll limit, one can ignore the three-point correlation functions of the scalar fields and take into account only the two-point functions
\beq
\langle
\delta\varphi^I_{\bk_1} \delta\varphi^J_{\bk_2} \rangle = (2
\pi)^3 \delta_{IJ} \delta^{(3)} (\bk_1 +
\bk_2) P(k_1), \qquad  P(k)\equiv\frac{2 \pi^2}{k^3} \P (k) , \qquad \P(k)  \equiv \frac{H_k^2}{4
\pi^2},
\eeq
The bispectrum (\ref{bispectrum}) can then be written as 
\beq
B_\zeta (\bk_1,\bk_2,\bk_3)=\frac65f_{\rm NL}\left[P(k_1)P(k_2)+P(k_2)P(k_3)+P(k_3)P(k_1)\right].
\eeq
with the non-linearity parameter
\beq
\label{f_NL_gen}
\frac{6}{5}f_{\rm NL} = 
  \frac{
N_{I} N_{J} N^{IJ}}{( N_{K}N^{K})^2}.
\eeq
 
The present observational constraints on the non-linearity parameter $f_{\rm NL}$, 
based on the  WMAP 5yr data, are \cite{WMAP5,Smith:2009jr}
\beq
-4< f^{\rm loc}_{\rm NL}<80 \quad (95\% {\rm CL})
\eeq
for the local type of non-Gaussianity considered here. The Planck satellite, which has just been launched, is expected to reach $f_{\rm NL}\sim 5$.

Using the Taylor expansion up to third order, one can compute in a similar way the trispectrum, i.e. the Fourier transform of the connected four-point function defined by
\beq
\langle \zeta_{\bk_1} \zeta_{\bk_2} \zeta_{\bk_3} \zeta_{\bk_4} \rangle_{c} \equiv (2 \pi)^3
\delta^{(3)}(\sum_i \bk_i) 
T_\zeta (\bk_1, \bk_2, \bk_3, \bk_4)\, .
\eeq
With the same assumptions as above, the trispectrum can be written in the form~\cite{Byrnes:2006vq}
\beq
\label{trispectrum}
T_\zeta (\bk_1, \bk_2, \bk_3, \bk_4)=\tau_{\rm NL}\left[P(k_{13})P (k_3)P(k_4)+ 11 \ {\rm perms}\right]
+\frac{54}{25} g_{\rm NL}\left[P(k_2)P(k_3)P(k_4)+3\ {\rm perms}\right],
\eeq
with 
 \beq
  \tau_{\rm NL}= \frac{N_{IJ}N^{IK}N^JN_K}{(N_LN^L)^3}, \qquad
  g_{\rm NL}=\frac{25}{54}\frac{N_{IJK}N^IN^J N^K}{(N_LN^L)^3}
  \eeq
and where $k_{13}\equiv\left|\bf {k}_1+\bf {k}_3\right|$.

After this general introduction, let us consider our particular model where the number of e-folds 
contains  two separate contributions, so that the Taylor  expansion up to third order is given by
\beq
\zeta=\delta N= \frac{d N_{\rm slow-roll}}{d\phi}\delta\phi+\dots+ \Delta N_{,\mod}\delta\mod+\frac12\Delta N_{,\mod\mod}\delta\mod^2
+\frac16\Delta N_{,\mod\mod\mod}\delta\mod^3
\eeq
where we have ignored the second and third derivatives with respect to the inflaton, which give negligible non-Gaussianities.
  According to (\ref{f_NL_gen}), the corresponding 
non-linearity parameter is given by
\beq
\label{f_NL}
\frac{6}{5}f_{\rm NL} = 
  \frac{
(\Delta N_{,\mod})^2 \Delta N_{,\mod\mod}}{\left({N^{\rm sr}_{,\phi}}^2+(\Delta N_{,\mod})^2\right)^2}
=\left(\frac{\P_\zeta^{\rm trapping}}{\P_\zeta}\right)^2\frac{\Delta N_{,\mod\mod}}{(\Delta N_{,\mod})^2}
=\X^2\, \frac{\Delta N_{,\mod\mod}}{(\Delta N_{,\mod})^2}
.
\eeq
Similarly, the coefficients of the trispectrum (\ref{trispectrum}) are 
\beq
\label{g_NL}
\tau_{\rm NL}
= \frac{(\Delta N_{,\mod\mod})^2}{(\Delta N_{,\mod})^4}\ \X^3=\frac{36}{25\X}f_{\rm NL}^2\,, \qquad
g_{\rm NL}=\frac{25}{54}\frac{\Delta N_{,\mod\mod\mod}}{(\Delta N_{,\mod})^3}\ \X^3
\eeq
So far, our  expressions (\ref{f_NL}-\ref{g_NL}) are quite similar  to those obtained in the context of mixed inflaton and curvaton models~\cite{Ichikawa:2008iq} or modulated reheating~\cite{Ichikawa:2008ne}.

Using now our explicit expressions (\ref{DeltaN2}) and (\ref{DeltaNprime}), one finds
 \begin{eqnarray}\label{long}
 \frac{\Delta N_{,\mod\mod}}{(\Delta N_{,\mod})^2} &=&
 \frac{3}{e\, \beta\, \left(5\,M_P
   \sqrt{2\epsilon_*}\,\lambda'+\left(\eta_*-3\, \epsilon_*\right)\,\lambda\,\phi_*'\right)^2}
   \times\nonumber\\
&&\times 
   \left[\left(21 \epsilon_*^2-8\eta_*\epsilon_*-\eta_*^2+2\,\sqrt{2\epsilon_*}\,M_P\,\eta'\right) \lambda^2 \phi_* '{}^2+10\,\epsilon_*\left(3\lambda'{}^2+2\,\lambda\,\lambda''\right)\,M_P^2+\right.\nonumber\\
&&+\left.2\,\sqrt{2\epsilon_*}\left(\eta_*-3\epsilon_*\right)\,\left(\lambda^2\,\phi_*''+5\,\lambda\,\lambda'\,\phi_*'\right)M_P
   \right]\, ,  \end{eqnarray}
   which we have expressed in terms of $\phi_*(\sigma)$ (rather than  in terms of $m(\sigma)$) for compactness, and
where we have used eq.~(\ref{beta}).
 Below, we specialize this expression  for the two particular cases which we have discussed in the previous section.

\subsection{Modulaton-independent coupling}

In the  case $\lambda=\lambda_0,\,m=g\mod$, we find from (\ref{f_NL}) and (\ref{long})
\begin{eqnarray}
f_{NL}&=&\frac{5}{2\, e\,\beta}\ \X^2\, \B_*\, , \qquad 
\B_*\equiv 
\frac{21\, \epsilon_*^2-8\, \eta_* \, \epsilon_* +2 \, M_P
 \,  \eta_*' \,\sqrt{2\epsilon_*}-\eta_*^2}{ (\eta_* -3\, \epsilon_*)^2}\,.
\end{eqnarray}
The term $\B_*$ in the above formula is typically of order one, since both the numerator and the denominator of the above formula are proportional to the square of the slow-roll parameters. For instance, chaotic inflation corresponds to $\eta=\epsilon$, $M_P\,\eta'=-\eta\sqrt{2\,\epsilon}$, and thus $\B_*=2$. Even if trapping dominates the spectrum, i.e. $\X\simeq1$, the non-linearity parameter cannot be much bigger than unity. 
Note also that, if $\eta_*\gg \epsilon_*$, one finds $\B_*=-1$ and thus a small but negative $f_{NL}$.

\subsection{Modulaton-dependent coupling}

In the  case $\lambda=\mod/M$ and $m=g\mod$,  (i.e. $\phi_*'=0$, $\lambda''=0$), we find the remarkably simple result
\beq
f_{NL}=\frac{3}{2\, e\, \beta}\, \X^2.
\eeq
The parameter $f_{NL}$ in this case depends only on the braking parameter $\beta$ and on the relative contribution of the trapping in  the power spectrum. If the trapping dominates, a small braking index $\beta$ can lead to a significant  $f_{\rm NL}$: for example, $\beta=10^{-2}$ gives $f_{NL}\simeq 55$.  However, one must be aware that a small $\beta$ tends to reduce $\X$ as well. In the small $\beta$ limit, $\X$ is proportional to $\beta^{6/5}$ and $f_{\rm NL} $ thus scales like $\beta^{7/5}$.

One can easily generalize the above result for an arbitrary  coupling and one finds 
\beq
f_{\rm NL}=\frac{1}{2e \beta}\left(3+2\frac{\lambda\lambda''}{\lambda'^2}\right)\X^2.
\eeq

One can also go beyond the bispectrum and study the trispectrum. Substituting the third derivative of $\Delta N$ with respect to $\sigma$ in the second expression of (\ref{g_NL}) yields
\beq
 g_{\rm NL}=\frac{1}{2e^2\beta^2}\left[1+6\frac{\lambda\lambda''}{\lambda'^2}+4\frac{\lambda^2\lambda'''}{3\lambda'^3}\right]\, \X^3.
  \eeq
It is interesting to observe that, if the coupling is of the form $\lambda(\mod)=\left(\mod/M\right)^p$, then one gets
\beq
f_{\rm NL}=\frac{\X^2}{2\, e \, \beta}\left(5-\frac{2}{p}\right),\,\, \tau_{\rm NL}=\frac{9\, \X^3}{25\, e^2 \beta^2}\left(5-\frac{2}{p}\right)^2,\,\,
g_{\rm NL}=\frac{\X^3}{6\, e^2\beta^2}\left(5-\frac{2}{p}\right)\left(5-\frac{4}{p}\right)\, .
\eeq
In this case there is a simple relation between the two coefficients $\tau_{\rm NL}$ and $g_{\rm NL}$, which depends only on the power $p$,
\beq
\tau_{\rm NL}=\frac{54 \, (5p-2)}{25 \, (5p-4)}\, g_{\rm NL},
\eeq
while $\X$ can be determined from $\tau_{\rm NL}$ and $f_{\rm NL}$ since $\X=(36/25)f^2_{\rm NL}/\tau_{\rm NL}$.

\section{Conclusions}

We have shown that modulated trapping, due to resonant particle production, provides a new mechanism for the  
conversion of  isocurvature into adiabatic perturbations. 
Although we have discussed mainly the trapping of  the  {\it  inflaton},
it is worth emphasizing that the same mechanism would apply to any rolling  scalar field whose 
energy density contributes significantly to the total matter budget in the Universe.

In the present work, we have computed the  perturbations generated by the trapping effect and compared their contribution in the power spectrum with respect to the usual contribution from the inflaton fluctuations. We have also computed the non-Gaussianities generated by the trapping effect. We have found that the non-linearity parameter $f_{\rm NL}$ of the bispectrum is proportional to the square of the fraction $\X$ 
of the power spectrum due to trapping, while the parameters $\tau_{\rm NL}$ and $g_{\rm NL}$ of the trispectrum are proportional to $\X^3$. This means that the non-Gaussianities are suppressed if $\X$ is small. 
To explore whether modulated trapping could account for most of the observed perturbations, it is convenient
to distinguish two main cases. 

In the first case, where the modulaton field affects only indirectly the particle production, the trapping contribution in the observed spectrum of fluctuations is expected to be small, except in two types of scenarios: if trapping occurs at a Hubble scale which is much smaller than the Hubble scale when the perturbations were produced, i.e. $H_*\ll H_k$; or if trapping occurs in a slow-roll phase where the slope of the potential is much smaller than its curvature, i.e. $\epsilon_*\ll \eta_*$.  Moreover, in this first case, the parameter $f_{\rm NL}$ is at most of order unity (if $\X\simeq 1$).

More interesting is the second case where the coupling $\lambda$ depends directly on the modulaton. It is then easy to obtain a domination of the trapping contribution in the observed signal. Moreover,
the non-linearity parameters $f_{\rm NL}$, $\tau_{\rm NL}$ and $g_{\rm NL}$, are given by very simple expressions that are inversely proportional  to the braking parameter $\beta$ or to its square.

To conclude, modulated trapping can be an efficient mechanism to generate primordial cosmological perturbations, with the possibility to produce significant non-Gaussianities that  could be detected in the near future in the CMB observations. It would be worthwhile to study further this mechanism in specific models of inflation embedded in realistic particle physics setups.

\vspace{1cm}

{\bf \noindent Acknowledgements}

\smallskip

We thank Andi Ross for interesting discussions. The work of LS is partially supported by the U.S. National Science Foundation grant PHY-0555304.


\vskip-1pc
\begin{thebibliography}{99}

\bibitem{Mukhanov:2005sc}
  V.~Mukhanov,
  ``Physical foundations of cosmology,''
{\it  Cambridge, UK: Univ. Pr. (2005) 421 p}

\bibitem{modulated}
  G.~Dvali, A.~Gruzinov and M.~Zaldarriaga,
  Phys.\ Rev.\  D {\bf 69}, 023505 (2004)
  [arXiv:astro-ph/0303591];
  L.~Kofman,
  arXiv:astro-ph/0303614.

\bibitem{curvaton}
  K.~Enqvist and M.~S.~Sloth,
  Nucl.\ Phys.\  B {\bf 626}, 395 (2002)
  [arXiv:hep-ph/0109214];
  D.~H.~Lyth and D.~Wands,
  Phys.\ Lett.\  B {\bf 524}, 5 (2002)
  [arXiv:hep-ph/0110002];
  T.~Moroi and T.~Takahashi,
  Phys.\ Lett.\  B {\bf 522}, 215 (2001)
  [Erratum-ibid.\  B {\bf 539}, 303 (2002)]
  [arXiv:hep-ph/0110096].
  
 

\bibitem{Kofman:1997yn}
  L.~Kofman, A.~D.~Linde and A.~A.~Starobinsky,
  Phys.\ Rev.\  D {\bf 56}, 3258 (1997)
  [arXiv:hep-ph/9704452].
  
\bibitem{Dolgov:1989us}
  A.~D.~Dolgov and D.~P.~Kirilova,
  Sov.\ J.\ Nucl.\ Phys.\  {\bf 51}, 172 (1990)
  [Yad.\ Fiz.\  {\bf 51}, 273 (1990)];
  J.~H.~Traschen and R.~H.~Brandenberger,
  ``Particle production during out-of-equilibrium phase transitions,''
  Phys.\ Rev.\  D {\bf 42}, 2491 (1990).


\bibitem{ckrt}
D.~J.~H.~Chung, E.~W.~Kolb, A.~Riotto and I.~I.~Tkachev,
   Phys.\ Rev.\  D {\bf 62}, 043508 (2000) [arXiv:hep-ph/9910437].
   
\bibitem{Elgaroy:2003hp}
  O.~Elgaroy, S.~Hannestad and T.~Haugboelle,
  JCAP {\bf 0309}, 008 (2003)
  [arXiv:astro-ph/0306229].

\bibitem{Romano:2008rr}
  A.~E.~Romano and M.~Sasaki,
  Phys.\ Rev.\  D {\bf 78}, 103522 (2008)
  [arXiv:0809.5142 [gr-qc]].
  
\bibitem{Kofman:2004yc}
  L.~Kofman, A.~Linde, X.~Liu, A.~Maloney, L.~McAllister and E.~Silverstein,
  JHEP {\bf 0405}, 030 (2004)
  [arXiv:hep-th/0403001].
  
\bibitem{Green:2009ds}
  D.~Green, B.~Horn, L.~Senatore and E.~Silverstein,
  arXiv:0902.1006 [hep-th].

\bibitem{Kohri:2009ac}
  K.~Kohri, D.~H.~Lyth and C.~A.~Valenzuela-Toledo,
  arXiv:0904.0793 [hep-ph].

\bibitem{Matsuda:2007tr}
  T.~Matsuda,
  JHEP {\bf 0707}, 035 (2007)
  [arXiv:0707.0543 [hep-th]].

\bibitem{Langlois:2004px}
  D.~Langlois and F.~Vernizzi,
  JCAP {\bf 0501}, 002 (2005)
  [arXiv:astro-ph/0409684].
  
\bibitem{Barnaby:2009mc}
  N.~Barnaby, Z.~Huang, L.~Kofman and D.~Pogosyan,
  arXiv:0902.0615 [hep-th].

\bibitem{deltaN}
  A.~A.~Starobinsky,
  JETP Lett.\  {\bf 42}, 152 (1985)
  [Pisma Zh.\ Eksp.\ Teor.\ Fiz.\  {\bf 42}, 124 (1985)];
  M.~Sasaki and E.~D.~Stewart,
  Prog.\ Theor.\ Phys.\  {\bf 95}, 71 (1996)
  [arXiv:astro-ph/9507001];
  M.~Sasaki and T.~Tanaka,
  Prog.\ Theor.\ Phys.\  {\bf 99}, 763 (1998)
  [arXiv:gr-qc/9801017].
  
\bibitem{Lyth:2004gb}
  D.~H.~Lyth, K.~A.~Malik and M.~Sasaki,
  JCAP {\bf 0505}, 004 (2005)
  [arXiv:astro-ph/0411220].
  
\bibitem{Langlois:2005ii}
  D.~Langlois and F.~Vernizzi,
  Phys.\ Rev.\ Lett.\  {\bf 95}, 091303 (2005)
  [arXiv:astro-ph/0503416].
  
\bibitem{Langlois:2005qp}
  D.~Langlois and F.~Vernizzi,
  Phys.\ Rev.\  D {\bf 72}, 103501 (2005)
  [arXiv:astro-ph/0509078]. 
  
\bibitem{Langlois:2004nn}
  D.~Langlois and F.~Vernizzi,
  Phys.\ Rev.\  D {\bf 70}, 063522 (2004)
  [arXiv:astro-ph/0403258].


\bibitem{Adams:1997de}
  J.~A.~Adams, G.~G.~Ross and S.~Sarkar,
  Nucl.\ Phys.\  B {\bf 503}, 405 (1997)
  [arXiv:hep-ph/9704286].

\bibitem{Burgess:2005sb}
  C.~P.~Burgess, R.~Easther, A.~Mazumdar, D.~F.~Mota and T.~Multamaki,
  JHEP {\bf 0505}, 067 (2005)
  [arXiv:hep-th/0501125].

\bibitem{Lyth:2005fi}
  D.~H.~Lyth and Y.~Rodriguez,
  Phys.\ Rev.\ Lett.\  {\bf 95}, 121302 (2005)
  [arXiv:astro-ph/0504045].
  
\bibitem{WMAP5}
  E.~Komatsu {\it et al.}  [WMAP Collaboration],
  Astrophys.\ J.\ Suppl.\  {\bf 180}, 330 (2009)
  [arXiv:0803.0547 [astro-ph]].

\bibitem{Smith:2009jr}
  K.~M.~Smith, L.~Senatore and M.~Zaldarriaga,
  arXiv:0901.2572 [astro-ph].

\bibitem{Byrnes:2006vq}
  C.~T.~Byrnes, M.~Sasaki and D.~Wands,
  Phys.\ Rev.\  D {\bf 74}, 123519 (2006)
  [arXiv:astro-ph/0611075].
  
\bibitem{Ichikawa:2008iq}
  K.~Ichikawa, T.~Suyama, T.~Takahashi and M.~Yamaguchi,
  Phys.\ Rev.\  D {\bf 78}, 023513 (2008)
  [arXiv:0802.4138 [astro-ph]].

\bibitem{Ichikawa:2008ne}
  K.~Ichikawa, T.~Suyama, T.~Takahashi and M.~Yamaguchi,
  Phys.\ Rev.\  D {\bf 78}, 063545 (2008)
  [arXiv:0807.3988 [astro-ph]].


  
\end{thebibliography}
\end{document}